\documentclass[a4paper]{article}
\usepackage{listings}
\usepackage{xcolor}
\usepackage{multirow}
\usepackage{jheppub}  
\usepackage{dcolumn}
\usepackage[english]{babel}
\usepackage[utf8x]{inputenc}
\usepackage[T1]{fontenc}
\usepackage{palatino}
\usepackage{amsmath}
\usepackage{afterpage}
\usepackage{graphicx, subcaption}
\usepackage[colorinlistoftodos]{todonotes}
\usepackage[colorlinks=true]{hyperref}
\usepackage{makeidx}
 
\usepackage[section]{placeins}
 
\makeindex
\begin{document}

\vspace*{1.2cm}

\thispagestyle{empty}
\begin{center}
{\LARGE \bf Parton distribution functions and intrinsic \\ \vspace{0.5cm} charm at LHCb}

\par\vspace*{7mm}\par

{

\bigskip

\large \bf Cristina Sánchez Gras} \\
{\large on behalf of the LHCb collaboration}

\bigskip


{Nikhef National Institute for Subatomic Physics, Amsterdam, The Netherlands}

\bigskip

{\it Presented at the Low-$x$ Workshop, Elba Island, Italy, September 27--October 1 2021}

\vspace*{15mm}

\end{center}
\vspace*{1mm}

\begin{abstract}
	AAt LHCb, proton parton distribution functions (PDFs) can be studied in a unique phase space complementary to that accessible by ATLAS and CMS, corresponding to low and high values of Bjorken-$x$. The measurements of vector boson production in the forward region, with and without an associated jet, are presented. These measurements can be used to constrain the proton PDFs, and in particular, the production of a $Z$ boson in association with a $c$-jet can be studied to measure the intrinsic charm content of the proton.
\end{abstract}
 
\section{Introduction}
Initially designed for the study of forward beauty and charm physics, the LHCb detector has the pseudorapidity ($\eta$) coverage $2 < \eta < 4.5$~\cite{lhcbdetector}. Its remarkable vertex reconstruction and particle identification performance, together with its high momentum resolution, have now established it as a general purpose detector. The forward coverage allows LHCb to reach \mbox{Bjorken-$x$} values (where $x$ is the fraction of momentum carried away by a parton) complementary to those accessible by other general purpose detectors, such as CMS~\cite{cms} and ATLAS~\cite{atlas}. This allows to probe proton parton distribution functions (PDFs) at very low- and high-$x$ values.

On the low-$x$ side, central exclusive production (CEP) of $J/\psi$ and $\psi(2S)$ mesons in proton-proton ($pp$) collisions can probe the gluon PDF down to $x \sim 10^{-6}$~\cite{lhcbCEP}. For high values of $x > 0.1$, $Z$ boson production in association with charm jets can be used to determine the intrinsic charm content of the proton~\cite{lhcbIC}. The most recent LHCb results for these two cases are presented in this note.

\section{\boldmath Central exclusive production of $J/\psi$ and $\psi(2S)$ mesons in $pp$ collisions}
The diffractive process \mbox{$pp \rightarrow p + X + p$} in which two protons stay unscathed following their interaction is known as central exclusive production (CEP). The proton interaction takes place via the exchange of colourless objects. In the case of vector meson production, the exchange of a photon and a pomeron receives the name of photoproduction. The cross-section for photoproduction is proportional to the square of the gluon PDF (at leading order) in perturbative quantumchromodynamics (QCD). In $pp$ collisions with the LHCb coverage, the gluon PDF can then be proved at very low-$x$ values of $x \sim 10^{-6}$.

This process has a very distinctive signature: low final state multiplicity, as only the muons that follow the meson decay are present in the detector, and large rapidity gaps (regions of no activity) around the dimuon system. The latter can be spoiled when one of the protons dissociates after the interaction, making it an inelastic CEP process. The proton remnants in that case are produced outside the $2 < \eta < 5$ coverage and escape detection. To veto these processes, three (two) scintillator stations are in place around the beam pipe upstream (downstream) from the interaction point, conforming HeRSCheL. The HeRSCheL (High Rapidity Shower Counters for LHCb) detector consists on five 60 cm x 60 cm stations equipped with four scintillating pads each~\cite{herschel}. It increases the LHCb coverage to $1.5 < \eta < 10$ and $-10 < \eta < 5, \ -3.5 < \eta < -1.5$ in the forward and backward regions, respectively. Charged particles produced when a proton dissociates trigger detection by the scintillating pads. A $\chi^2$-like figure of merit built with the activity registered at the HeRSCheL stations is used to veto these events. 

\begin{figure}[hb]
	\begin{center}
		\begin{tabular}{c}
			\includegraphics[width=0.49\textwidth]{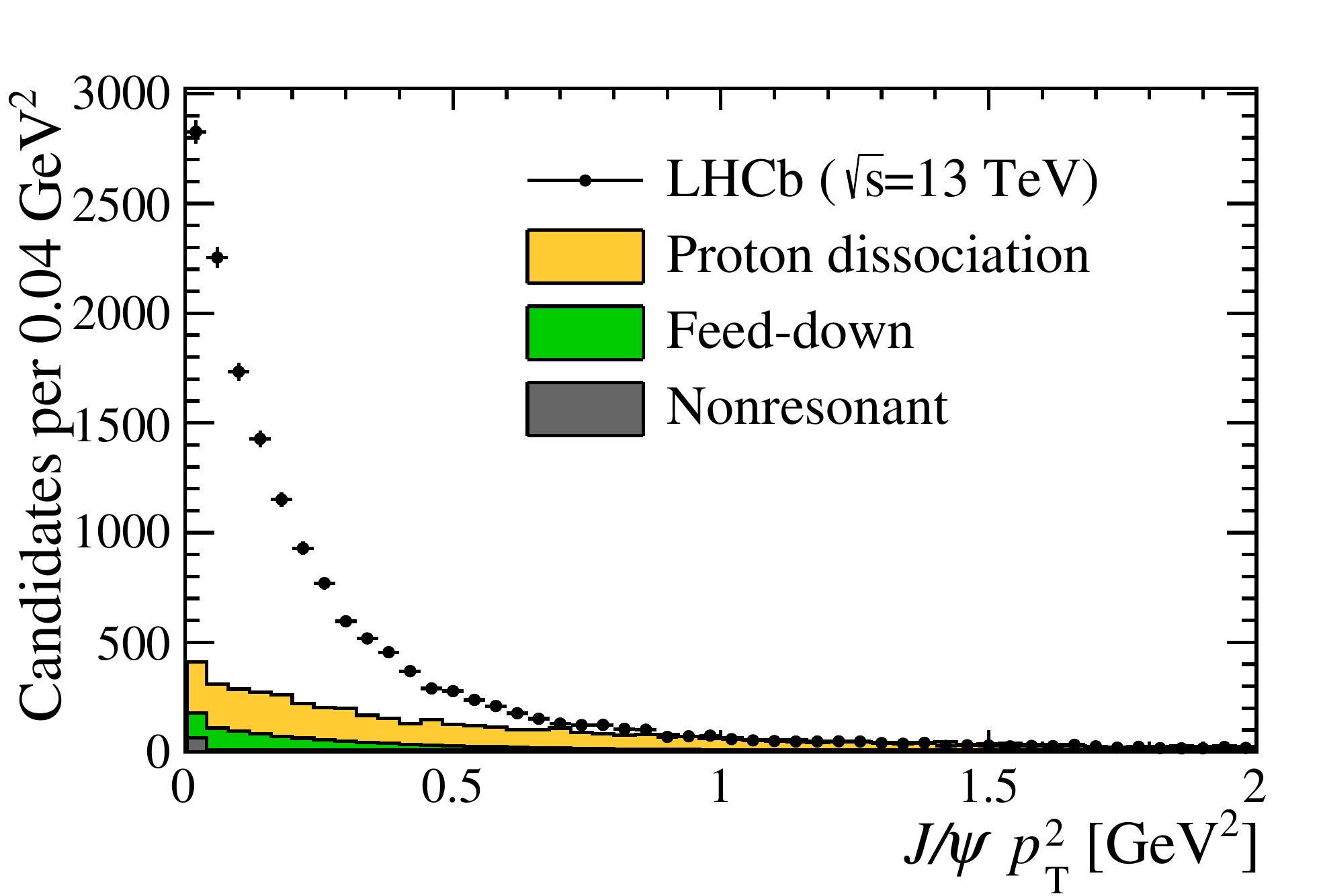}\hskip 0.01\textwidth
			\includegraphics[width=0.49\textwidth]{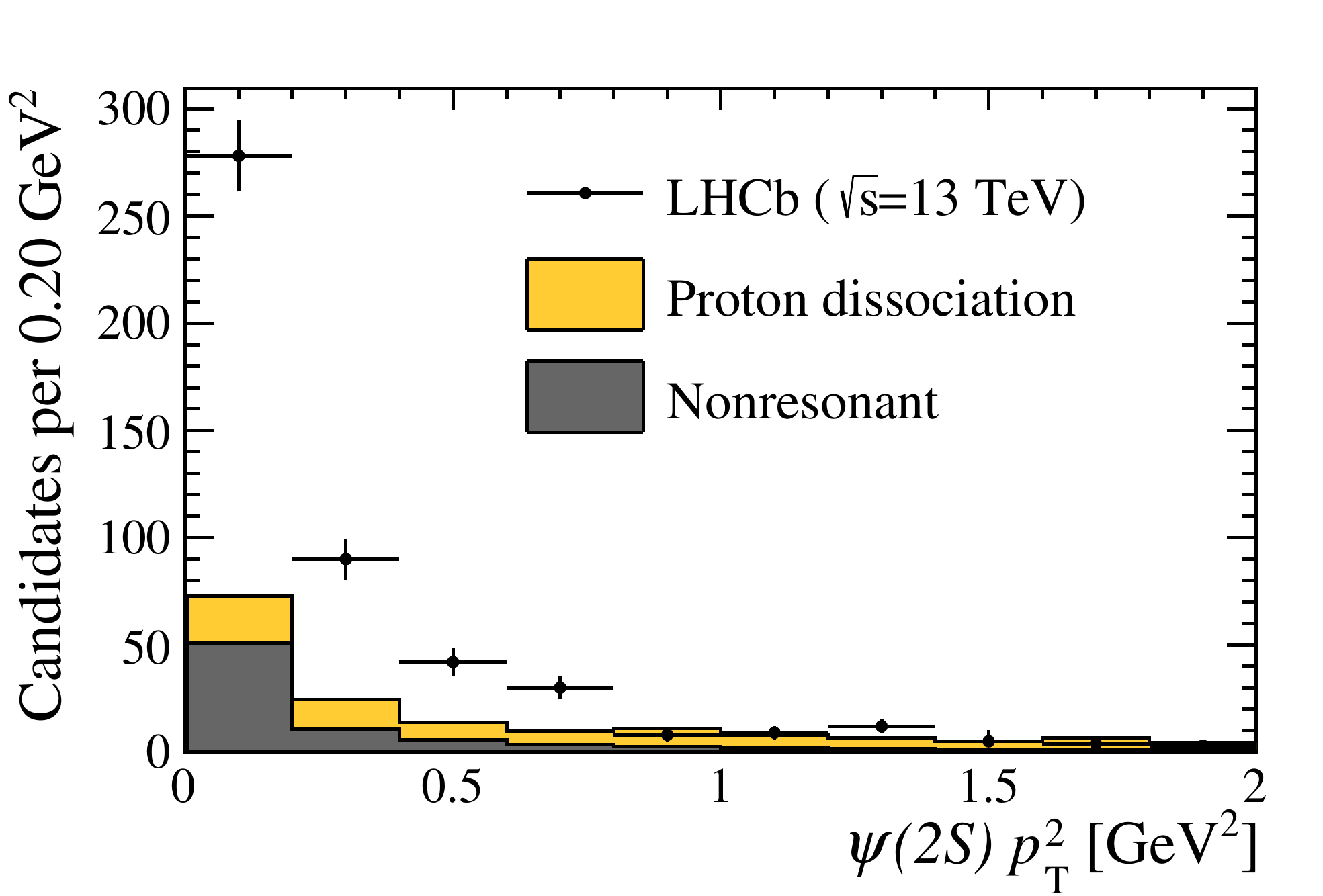}\\
		\end{tabular}
		\caption{Squared transverse momentum ($p_{\rm T}^2$) distribution for CEP $J/\psi \rightarrow \mu^+\mu^-$ (left) and \mbox{$\psi(2S) \rightarrow \mu^+\mu^-$} (right). The different backgrounds described in the text are indicated. }
		\label{fig:ptsqCEP}
	\end{center}
\end{figure}

CEP events are selected by requiring exactly two muon tracks and imposing a veto on the activity in HeRSCheL. The number of signal events is obtained by fitting the dimuon squared transverse momentum ($p_{\rm T}^2$) distribution. Following Regge theory, the cross-section for $J/\psi$ and $\psi(2S)$ CEP events follows \mbox{${\rm d \sigma}/{\rm d}y \sim \exp(-bp_{\rm T}^2)$}, with $b \sim 6$ (GeV$/c)^{-2}$. The background arising from the dissociation of one of the protons is parametrised in a sample with the HeRSCheL veto inverted. Aside from this, two more backgrounds need to be accounted for: nonresonant dimuon production and feed-down to $J/\psi$ from $\psi(2S)$ and $\chi_{c_{J}} \ (J=0,1,2)$. Nonresonant dimuon production takes place when both protons interact electromagnetically via photon-photon exchange. This background is measured by fitting the dimuon mass spectrum, and its $p_{\rm T}^2$ shape modelled from simulated samples. The $J/\psi$ feed-down background concerns $\chi_{c_{J}} \rightarrow J/\psi \gamma$ and $\psi(2S) \rightarrow J\psi X$ decays where only the $J/\psi$ is fully reconstructed. Its contribution is estimated combining data and simulation. The $J/\psi$ and $\psi(2S)$ $p_{\rm T}^2$ distributions are shown in Fig.~\ref{fig:ptsqCEP}. A fit to the background-subtracted shapes is used to determine the number of signal events.

\begin{figure}[t]
	\begin{center}
		\begin{tabular}{c}
			\includegraphics[width=0.49\textwidth]{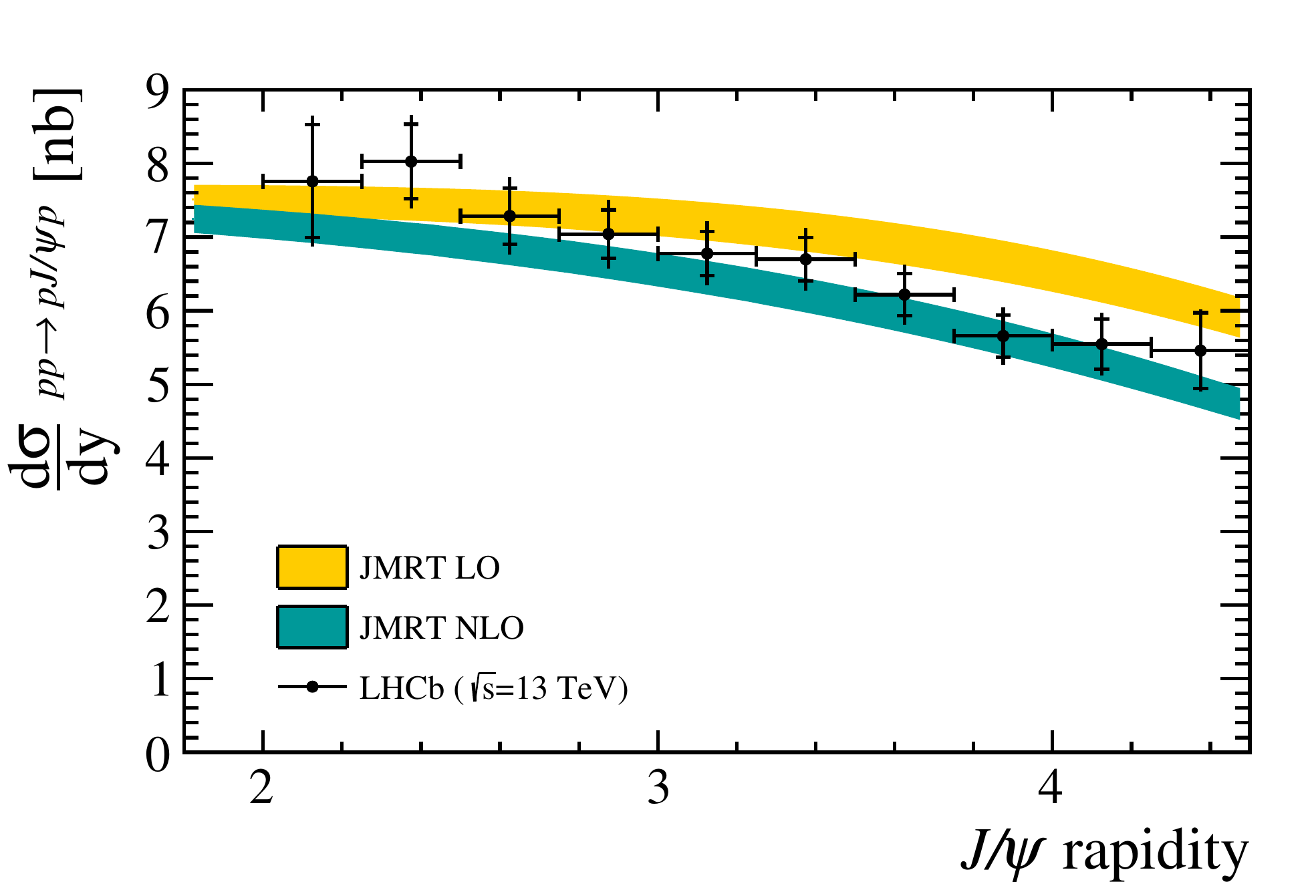}\hskip 0.01\textwidth
			\includegraphics[width=0.49\textwidth]{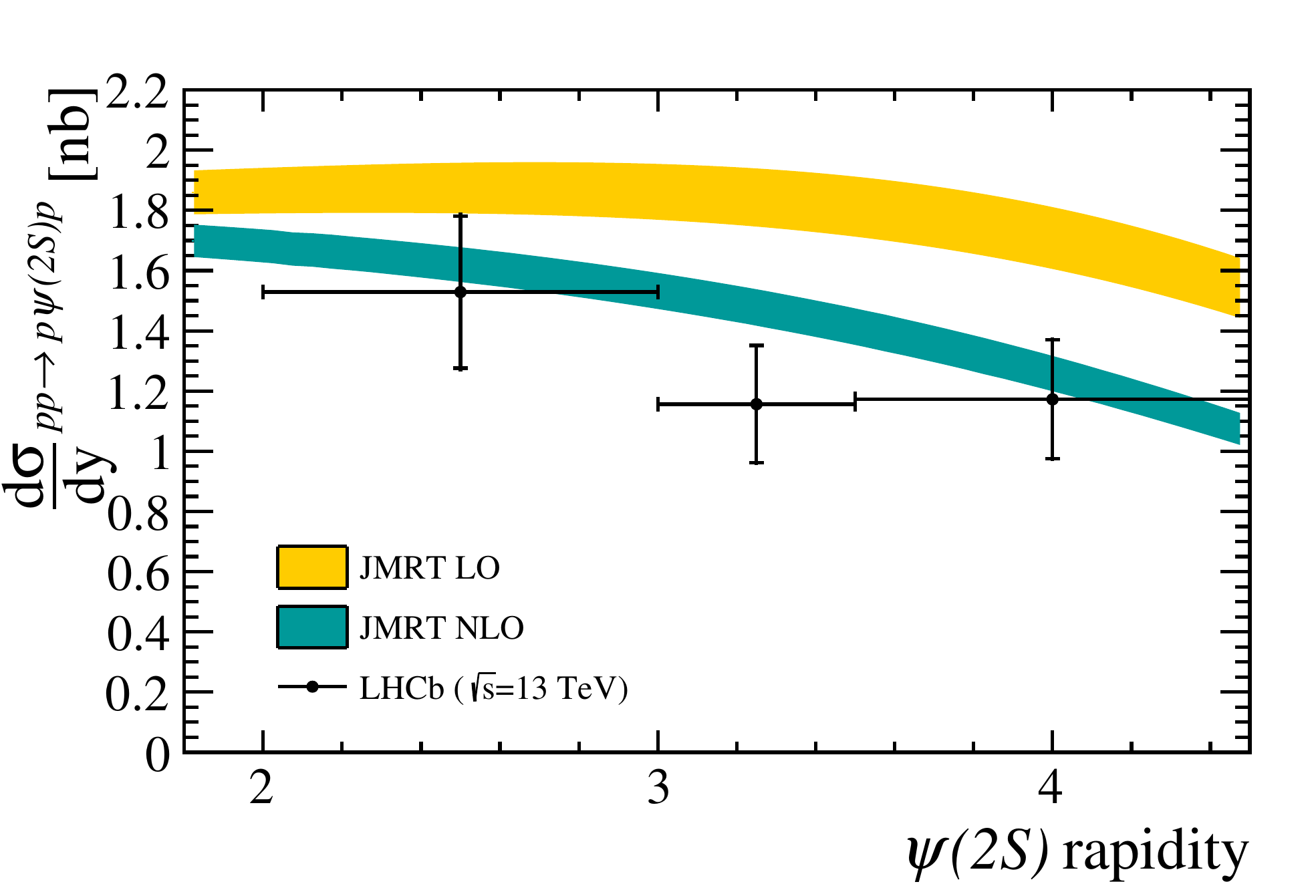}\\
		\end{tabular}
		\caption{Differential cross-section results for $J/\psi \rightarrow \mu^+\mu^-$ (left) and \mbox{$\psi(2S) \rightarrow \mu^+\mu^-$} (right) compared to LO and NLO JMRT theory descriptions~\cite{JMRTLO,JMRTNLO}. }
		\label{fig:resultCEP}
	\end{center}
	\begin{center}
		\begin{tabular}{c}
			\includegraphics[width=0.49\textwidth]{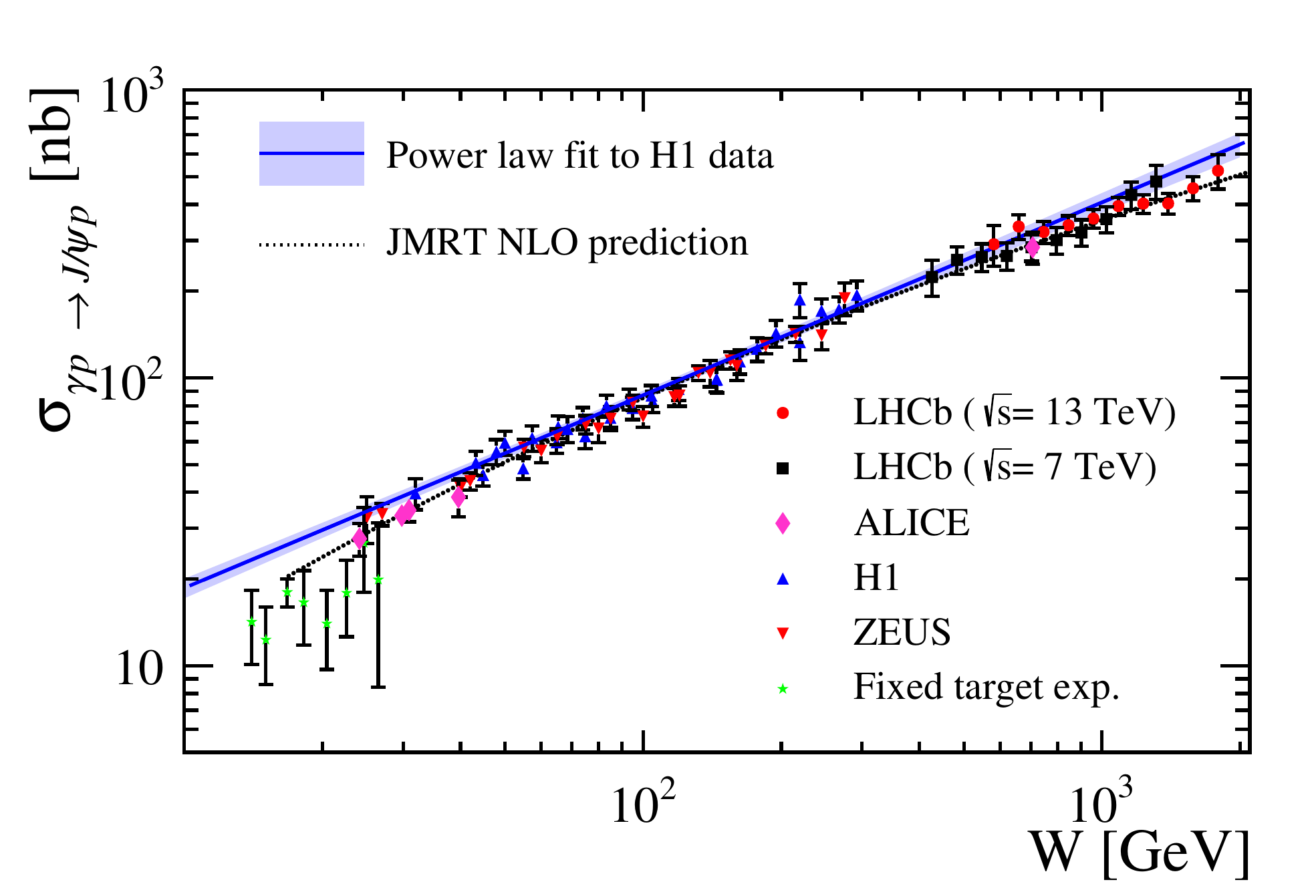}\hskip 0.01\textwidth
			\includegraphics[width=0.49\textwidth]{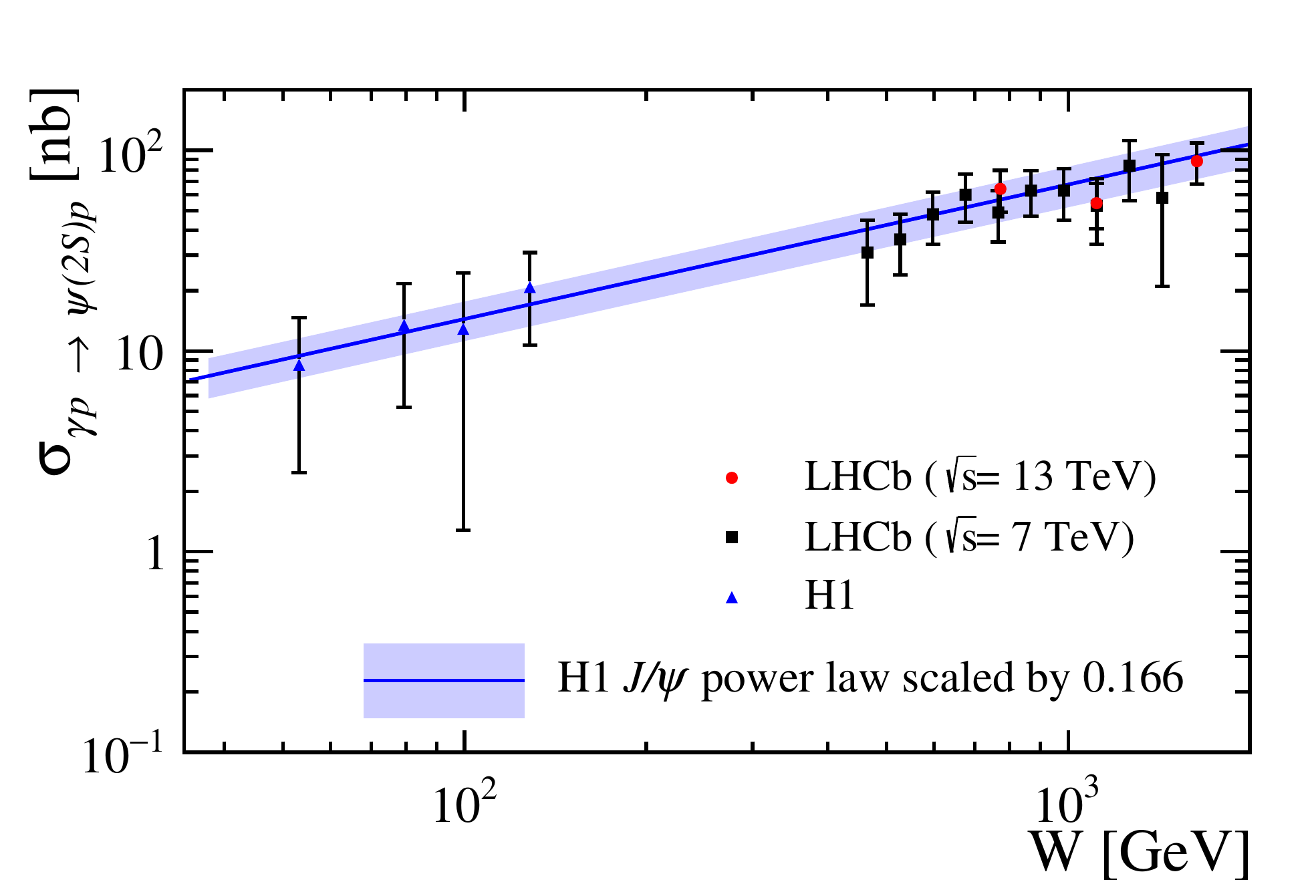}\\
		\end{tabular}
		\caption{Photoproduction cross-section results for $J/\psi$ (left) and $\psi(2S)$. The LHCb result at $\sqrt{s}=13 \ \mathrm{TeV}$ is shown together with the LHCb results at $\sqrt{s}=7 \ \mathrm{TeV}$~\cite{lhcb7tev} and those from the H1~\cite{H1param,H1psi2S} ZEUS~\cite{zeusCEP} and ALICE~\cite{aliceCEP} collaborations, and the results from the fixed target experiments E401~\cite{e401}, E516~\cite{e516} and E687~\cite{e687}.}
		\label{fig:photoprodCEP}
	\end{center}
\end{figure}

The result for the differential cross-section in rapidity bins is shown in Fig.~\ref{fig:resultCEP}, as well as the leading order (LO) and next-to-leading order (NLO) Jones-Martin-Ryskin-Teubner (JMRT) theory descriptions~\cite{JMRTLO,JMRTNLO}. In the $J/\psi$  case, the data is more in agreement with the NLO description, especially in the higher rapidity bins. For the $\psi(2S)$ the same trend is observed, but higher statistics are needed.

The measured cross-section per rapidity bins allow for the calculation of the photoproduction cross-section, $\sigma_{pp \rightarrow p\psi p}$, as:

\begin{equation}\label{eq:photoprod}
	\sigma_{pp \rightarrow p\psi p} = r(W_+)k_+ \frac{\mathrm{d}n}{\mathrm{d}k_+} \sigma_{\gamma p \rightarrow \psi p} (W_+) \ +  \ r(W_-)k_- \frac{\mathrm{d}n}{\mathrm{d}k_-} \sigma_{\gamma p \rightarrow \psi p} (W_-)\,,
\end{equation}
where $r(W_{\pm}$ is the gap survival factor, $k_{\pm} \equiv M_{\psi}/2e^{\pm |y|}$ is the photon energy, $\frac{\mathrm{d}n}{\mathrm{d}k_{\pm}}$ is the photon flux and $W_{\pm} = 2k_{\pm}\sqrt{s}$ is the photon-proton system invariant mass. The positive (negative) signs in Eq.~\ref{eq:photoprod} refer to the situation where the photon is emitted by the proton travelling parallel (antiparallel) to the LHCb beam axis. In LHCb, $W_+$ and $W_-$ contribute to the same rapidity bin and cannot be disentangled. However, given that only about a third of the data corresponds to $W_-$ and this low-energy contribution has been parametrised for the $J/\psi$ meson by the H1 collaboration~\cite{H1param}, their power-law parametrisation is used to fix it. This power-law is scaled by the ratio of the $\psi(2S)$ and $J/\psi$ cross-sections measured by H1~\cite{H1psi2S}. The estimated photoproduction cross-section is presented in Fig.~\ref{fig:photoprodCEP} and compared to the power-law H1 fit and their results~\cite{H1param,H1psi2S}, as well as to different results from other experiments. Also shown is the JMRT NLO theory description. In the case of the $J/\psi$, where more data is available, it is observed that the LHCb photoproduction cross-section values at $\sqrt{s} = 13 \ \mathrm{TeV}$ deviate from the power-law fit at higher rapidities and are in more agreement with the JMRT NLO description. More data is necessary to discern the behaviour of the $\psi(2S)$ photoproduction cross-section at high rapidities.

\section{\boldmath Intrinsic charm with $Z$ bosons produced in association with charm jets}
While the extrinsic charm ($c$) content of the proton (due to perturbative gluon radiation) has been widely established, several theory predictions suggest that the proton also contains charm intrinsically. This could take place in a sea-quark-like manner or as a valence-like $c$ quark, transforming the proton wave-function into $|uudc\bar{c}\rangle$, as predicted by Light Front QCD (LFQCD). Previous measurements have been performed at low-$Q^2$ \cite{measIC1,measIC2}. At such low energy, the theoretical treatment of hadronic nuclear effects is challenging, and it is difficult to understand the results as evidence or not of the proton intrinsic charm (IC) content. Nevertheless, global PDF analysis do not exclude it at the percent level \cite{percIC1,percIC2}.

\begin{figure}[ht]
	\begin{center}
		\includegraphics[width=0.49\textwidth]{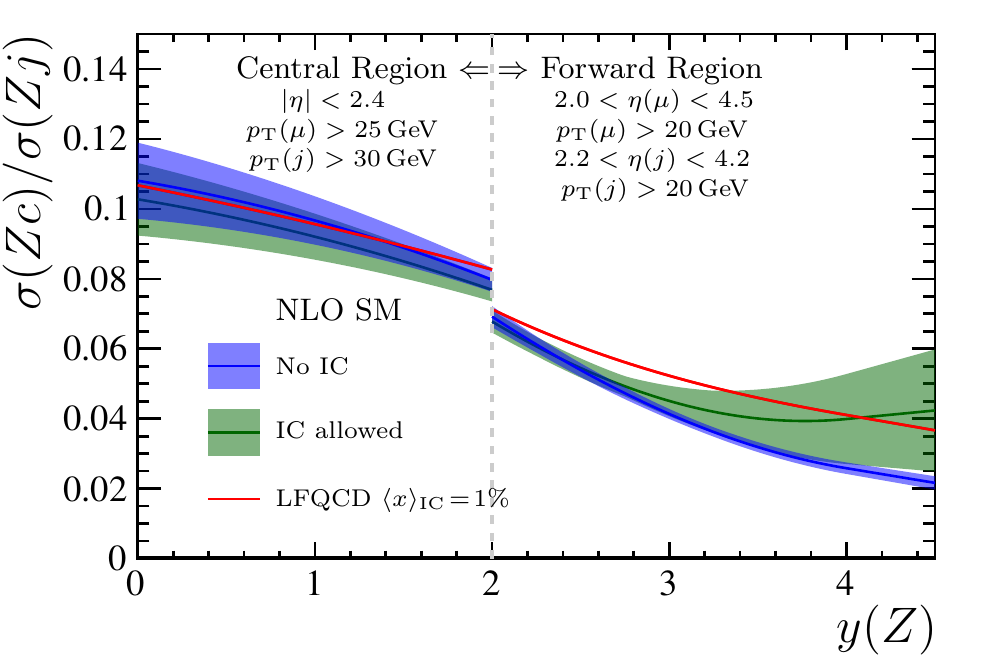}
		\caption{Theory predictions including and excluding intrinsic charm content in the proton for $\sigma(Zc)/\sigma(Zj)$~\cite{proposalIC}. The range $2 < y(Z) < 4.5$ corresponds to the LHCb forward region.}
		\label{fig:theoryIC}
	\end{center}
\end{figure}

A proposal was made to measure the ratio of $Z+c$ jets events to that of $Z+$jets, $\sigma(Zc)/\sigma(Zj)$ in the forward region~\cite{proposalIC}. Performing this measurement at high-$Q^2$ with $Z$ bosons at forward rapidities allows to access high Bjorken-$x$ values with $x>0.1$, where the hadronic and nuclear effects are negligible. Figure~\ref{fig:theoryIC} illustrates how the ratio $\sigma(Zc)/\sigma(Zj)$ at high $Z$ boson rapidities would allow to discriminate the intrinsic charm content of the proton.

\begin{figure}[ht]
	\begin{center}
		\begin{tabular}{c}
			\includegraphics[width=0.49\textwidth]{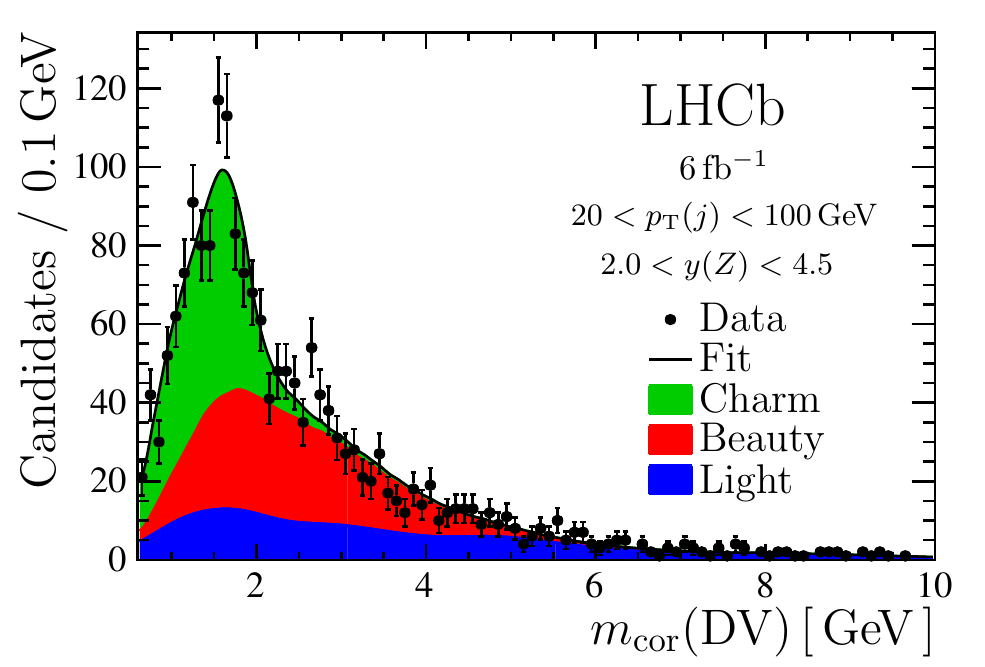}\hskip 0.01\textwidth
			\includegraphics[width=0.49\textwidth]{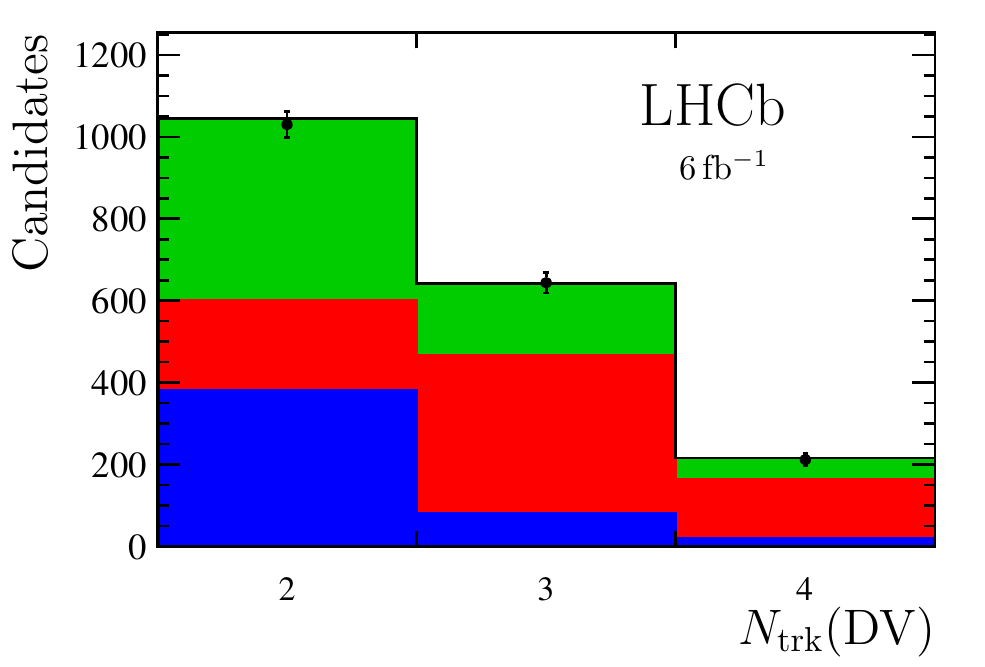}\\
		\end{tabular}
		\caption{Result of the two-dimensional fit for the corrected mass ($m_\mathrm{cor}$) and number of tracks ($N_\mathrm{trk}$) in the dispaced-vertex (DV). The contributions for charm, beauty and light jets are shown.}
		\label{fig:fitIC}
	\end{center}
\end{figure}

An integrated luminosity of 6 fb$^{-1}$ corresponding to the full proton-proton collision LHCb dataset at $\sqrt{s} = 13 \ \mathrm{TeV}$ is used \cite{lhcbIC}. Events with $Z \rightarrow \mu^+ \mu^-$ and at least one jet with transverse momentum  \mbox{$p_\mathrm{T} > 20 \ \mathrm{GeV/}c$} are selected. Charm jets are identified by using a displaced-vertex (DV) tagger in bins of $p_\mathrm{T}(\mathrm{jet}), y(Z)$. A two-dimensional fit to the corrected DV mass and the number of tracks in the DV is performed to identify the flavour of each jet in the selected events. The result of the fit is shown in Fig.~\ref{fig:fitIC}. The efficiency of tagging a jet as a charm jet is estimated in simulation and calibrated in data. The $Z_c$ and $Z_j$ yields in each $y(Z)$ bin are corrected for their selection efficiency and resolution effects at detection.

\begin{figure}[ht]
	\begin{center}
		\includegraphics[width=0.49\textwidth]{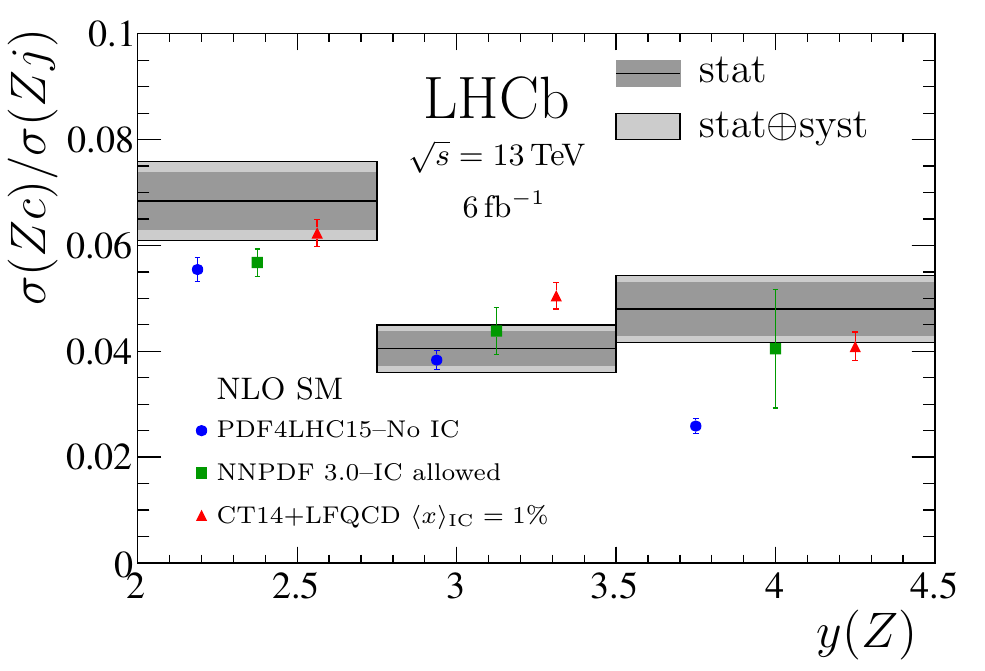}
		\caption{Results for $\sigma(Zc)/\sigma(Zj)$ compared to theory predictions allowing and excluding intrinsic charm content in the proton.}
		\label{fig:resultIC}
	\end{center}
\end{figure}

The measurement of a ratio results in most of the systematic uncertainties cancelling out. The dominant systematic uncertainty is related to the efficiency of identifying charm jets. A comparison of the measured $\sigma(Zc)/\sigma(Zj)$ values to different theory predictions is shown in Fig.~\ref{fig:resultIC}. The first two bins are compatible with both no IC and IC allowed content. The bin at higher $Z$ boson rapidity is consistent with proton IC models, and is about three standard deviations away from the prediction of no intrinsic charm content. The measurement is statistically limited and more data is needed to draw further conclusions. Moreover, these results need to be added to global PDF analyses for interpretation.

\section{Conclusions}
The LHCb detector can be used to perform precision QCD measurements, both in the low- and high-$x$ regions. The central exclusive production cross-section of $J/\psi$ allows to probe the region $x \sim 10^{-6}$. This measurement at $\sqrt{s} = 13 \ \mathrm{TeV}$ is in agreement with the JMRT NLO description, and further data is needed to observe if the same behaviour is present for $\psi(2S)$. The high-$x$ region provides access to large values $x > 0.1$, where the intrinsic charm of the content can be probed. While statistically limited, the first measurement of the proton intrinsic charm content in the forward region in proton-proton collisions has been performed.

\FloatBarrier
 

\end{document}